\documentclass[prc,superscriptaddress,unsortedaddress,twocolumn,showpacs,preprintnumbers,amsmath,amssymb]{revtex4}
\usepackage[dvips]{graphicx}
\usepackage{dcolumn}
\usepackage{amsmath}
\usepackage{times}
\begin{document}

\title{Continuum-discretized coupled-channels method for
four-body nuclear breakup in $^6$He+$^{12}$C scattering}

\author{T. Matsumoto}
\email[Electronic address: ]{taku2scp@mbox.nc.kyushu-u.ac.jp}
\affiliation{Department of Physics, Kyushu University, Fukuoka 812-8581,
Japan}

\author{E. Hiyama}
\affiliation{Department of Physics, Nara Women's University, Nara 630-8506, Japan}

\author{K. Ogata}
\affiliation{Department of Physics, Kyushu University, Fukuoka 812-8581,
Japan}

\author{Y. Iseri}
\affiliation{Department of Physics, Chiba-Keizai College, Todoroki-cho
4-3-30, Inage, Chiba 263-0021, Japan}

\author{M. Kamimura}
\affiliation{Department of Physics, Kyushu University, Fukuoka 812-8581,
Japan}

\author{S. Chiba}
\affiliation{Advanced Science Research Center,
Japan Atom Energy Research Institute (JAERI),
Tokai, Ibaraki 319-1195, Japan}

\author{M. Yahiro}
\affiliation{Department of Physics, Kyushu University, Fukuoka 812-8581,
Japan}

\date{\today}

\begin{abstract}
 We propose a fully quantum-mechanical method of treating four-body
 nuclear breakup processes in scattering of a projectile
 consisting of three constituents, by extending the
 continuum-discretized coupled-channels method. The three-body
 continuum states
 of the projectile are discretized by diagonalizing the internal
 Hamiltonian of the projectile with the Gaussian basis functions. For
 $^6$He+$^{12}$C scattering at 18 and 229.8 MeV, the validity of the
 method is tested by convergence of the elastic and breakup cross
 sections with respect to increasing the number of the basis
 functions. Effects of the four-body breakup and
 the Borromean structure of $^6$He on the elastic
 and total reaction cross sections are discussed.
\end{abstract}

\pacs{21.45.+v, 21.60.Gx, 24.10.Eq, 25.60.-t}

\maketitle

The study on neutron-halo nuclei
has become one of the central subjects in the unstable nuclear physics
since the discovery of such nuclei~\cite{Tanihata}.
In scattering of a two-neutron-halo nucleus
such as $^6$He and $^{11}$Li, the projectile easily breaks up into its
three constituents ($n$+$n$+core), indicating that the scattering
should be described as a four-body ($n$+$n$+core+target) reaction.
Then an accurate theory for treating such a four-body breakup is
highly desirable.

So far the eikonal and adiabatic calculations
were proposed and applied to $^6$He and $^{11}$Li scattering
around 50 MeV/nucleon~\cite{adiabatic1,adiabatic2,eikonal,exp6He2}.
Since these calculations are based on semi-classical approaches, they
work well at higher incident energies.
In fact, the elastic cross section of $^{6}$He+$^{12}$C scattering
at 229.8 MeV has recently  been measured~\cite{exp6He1} and successfully
analyzed by the eikonal calculation with the six-nucleon wave function of
$^6$He~\cite{BAI}. However, these approaches seem not to be applicable
for low-energy scattering such as $^{12}$C($^{6}$He,$^{6}$He)$^{12}$C at
3 MeV/nucleon~\cite{Milin} measured very recently.

In this rapid communication, we present a fully quantum-mechanical
method of treating four-body nuclear breakup. The method
is constructed by extending the
continuum-discretized coupled-channels method (CDCC)~\cite{cdcc} that
treats three-body breakup processes in scattering of the two-body
projectile. In CDCC, the total scattering wave function is expanded in
terms of bound and continuum states of the
projectile. The continuum states are classified by the linear ($k$)
and angular momenta, and they are truncated by setting an upper
limit to each quantum number. The $k$-continuum is then
divided into small bins and the
continuum states in each bin are averaged into a single state.
This procedure of discretization is called the average (Av)
method.
The $S$-matrix elements calculated with CDCC converge as the modelspace
is extended~\cite{cdcc}.
The converged CDCC solution is the unperturbed solution of the distorted
Faddeev equations, and corrections to the solution are negligible within
the region of space in which the reaction takes
place~\cite{CDCC-foundation}.

Also for four-body breakup processes in scattering of the three-body
projectile, CDCC has to prepare three-body bound and
discretized-continuum states of the projectile. Because of the
difficulty of preparing all the three-body states with the Av method,
CDCC so far analyzed $^6$He scattering within a limited model in which a
two-neutron pair is treated as a single particle, di-neutron
($^2n$)~\cite{dicdc}. However, the accuracy of the di-neutron model has
not been confirmed yet, because of the absence of fully quantum-mechanical
method of treating four-body breakup.

In our previous work~\cite{pscdc} on three-body breakup in
scattering of the two-body projectile,
we proposed a new method of discretization, called the pseudo-state
(PS) method. In the method, continuum states of the projectile are replaced
by discrete pseudo-states obtained by
diagonalizing the internal Hamiltonian of the projectile in a space
spanned by the $L^2$-type Gaussian basis functions.
The CDCC solution calculated by the PS method agrees with
that by the Av method which can be regarded as the exact solution.
Thus, a reasonable number of the Gaussian basis functions can
form an approximate complete set in a finite configuration space being
important for three-body breakup processes.
It is very likely that the approximate completeness persists also
in the case of four-body breakup processes. Actually, as shown latter,
we can see clear convergence of calculated elastic and breakup cross
sections with respect to increasing the number of the Gaussian basis
functions.
It should be noted that
the Gaussian basis functions are widely used
to solve bound-state problems of few-body systems~\cite{GEM},
since the use of the basis functions reduces numerical works much.
Thus, the four-body breakup processes can be analyzed properly by
CDCC with the PS method. We refer to this new method as
{\it four-body CDCC} and the usual CDCC for three-body breakup as
{\it three-body CDCC}.

The first application of four-body CDCC thus designed is made for
$^6$He+$^{12}$C scattering at 18 and 229.8 MeV, where the projectile
has the Borromean structure and then easily breaks up
into two nucleons and $^4$He. In
these scattering processes, the incident energies $E_{\rm in}$
are much higher than the Coulomb barrier energy
($\sim 3$ MeV), so only nuclear breakup processes become
significant. We thus concentrate our application on nuclear breakup.
The calculated elastic cross sections well reproduce experimental data
at both $E_{\rm in}$.
Moreover, effects of the four-body breakup and the Borromean structure of
$^6$He on the elastic and total reaction
cross sections are discussed in the case of $E_{\rm in}=$18 MeV.

We assume that $^6$He+$^{12}$C scattering is described as a
four-body system, $n$+$n$+$^4$He+$^{12}$C. Then, the
Schr\"{o}dinger equation can be written as
\begin{eqnarray}
 \left[
  K_{R}+\sum_{i\in{\rm P}}\sum_{j\in{\rm T}}v_{ij}+V_{\rm C}(R)
  +H_6-E
\right]
 \Psi(\xi,{\bf R})=0,\label{four-sch}
\end{eqnarray}
where ${\bf R}$ and $\xi$ are, respectively, the coordinate of
the center-of-mass of $^{6}$He relative to $^{12}$C and the internal
coordinates of $^6$He; $K_{R}$ is the kinetic energy
associated with ${\bf R}$. Here, $H_6$ is the internal Hamiltonian of
$^6$He-projectile, and $E$ is the sum of $E_{\rm{in}}$
and the ground state energy of $^6$He.
The $v_{ij}$ represent two-body nuclear
interactions working between $^6$He-projectile (P) and $^{12}$C-target (T).
Meanwhile, the Coulomb
potential $V_{\rm C}$ is treated approximately as a function of $R$
only, i.e., we neglect Coulomb breakup processes.

\begin{figure}[htbp]
\begin{center}
 \includegraphics[width=0.38\textwidth,clip]{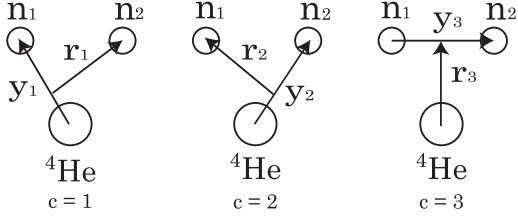}
 \caption{Jacobian coordinates of three rearrangement
 channels ($c=1\mbox{--}3$) adopted for the $n$+$n$+$^4$He model of $^6$He
 structure.}
 \label{Jacobi}
\end{center}
\end{figure}

The four-body wave function $\Psi^{JM}$, where $J$ is the total angular
momentum of the four-body system and $M$ is its projection on the $z$-axis,
is expanded in terms of a finite number of the internal wave functions
$\Phi_{\gamma}$ of $^6$He-projectile:
\begin{equation}
\Psi^{JM}(\xi,{\bf R})
=\sum_{nI,L}
\chi_{nI,L}^J (P_{nI},R)/R
\;{\cal Y}_{nI,L}^{JM},
\end{equation}
where ${\cal Y}_{nI,L}^{JM}=[\Phi_{nI}(\xi)\otimes i^L Y_L(\Omega_R)]_{JM}$.
The $\gamma$ stands for the set of $(n,I,m)$, where
$I$ is the total spin of $^6$He and $m$ is its projection
on the $z$-axis, and $n$ stands for the $n$ th eigenstate with positive
energy. The ground state of $^6$He, which is the only bound state of
$^6$He, is denoted by $\gamma_0\equiv (0,I_0,m_0)$.
The $\Phi_{\gamma}$ satisfies
$H_6 \Phi_{\gamma}=\epsilon_{nI} \Phi_{\gamma}$
and the expansion-coefficient $\chi_{nI,L}^J$ in Eq.~(\ref{four-sch})
represents the relative motion between the projectile and the target;
$L$ is the orbital angular momentum regarding ${\bf R}$.
The relative momentum $P_{nI}$ is determined by the conservation
of the total energy:
$E=P_{nI}^{2}/2\mu+\epsilon_{nI}$, with $\mu$ the reduced mass
between the projectile and the target. Multiplying Eq.~(\ref{four-sch})
by ${\cal Y}_{n'I',L'}^{JM}$ from left,
one can obtain a set of coupled
differential equations for $\chi_{nI,L}^J$, called CDCC equation;
it should be noted that the CDCC equation for the four-body system is
formally equal to that for the three-body
system. Solving the CDCC equation under the appropriate asymptotic boundary
condition~\cite{cdcc,Piya},
we can obtain the elastic and discrete breakup $S$-matrix
elements. Details of the formalism of CDCC are shown in Ref.~\cite{cdcc}.

\begin{table}[bp]
 \caption{The maximum internal angular momenta and the Gaussian
 range parameters for each Jacobian coordinate.}
 \begin{tabular}{ccccccccc}
  \hline
  $c$&$I^\pi$&$\lambda_{\rm max}$&$\ell_{\rm max}$&$\Lambda_{\rm max}$&
  $\bar{y}_1$ [fm]&$\bar{y}_{\rm max}$ [fm]
  &$\bar{r}_1$ [fm]&$\bar{r}_{\rm max}$ [fm]\\ \hline\hline
  3&$0^+$ & 1&1&1&  0.1&       10.0&   0.5&  10.0 \\
  1,2&$0^+$ & 1&1&1&  0.5&       10.0&   0.5&  10.0 \\ \hline
  3&$2^+$ & 2&2&2&  0.5&       10.0&   0.5&  8.0 \\
  1,2&$2^+$ & 1&1&2&  0.5&       10.0&   0.5&  8.0 \\ \hline
 \end{tabular}
 \label{tab1}
\end{table}
In the Gaussian expansion method (GEM)~\cite{GEM},
each $\Phi_{\gamma}$ is given by
\begin{equation}
\Phi_{\gamma}(\xi)=\textstyle{\sum_{c=1}^3} \psi^{(c)}_{\gamma}(\xi),
\label{phig}
\end{equation}
where $c$ denotes the set of Jacobian coordinates shown in Fig.~\ref{Jacobi}.
We here take the angular-momentum coupling scheme as
${\bf I}={\bf \Lambda}+{\bf S}$, where $\Lambda$
and $S$ are the total orbital-angular-momentum and the total intrinsic-spin
of $^6$He, respectively. Then $\psi^{(c)}_{\gamma}$ has the following form:
\begin{equation}
\psi^{(c)}_{\gamma}(\xi)=\varphi^{(\alpha)}\sum_{\Lambda,S}
\left[
\phi_{n\Lambda}^{(c)}({\bf y}_c,{\bf r}_c)
\otimes
\left[
\eta_{\frac{1}{2}}^{(n_1)}\otimes\eta_{\frac{1}{2}}^{(n_2)}
\right]_{S}
\right]_{Im},
\label{angcp}
\end{equation}
where $\eta_{1/2}$ is the spin wave function of each valence neutron
($n_1$ or $n_2$) and
$^4$He has been treated as an inert core
with the $(0s)^4$ internal configuration,
$\varphi^{(\alpha)}$.
The definition of $({\bf y}_c,{\bf r}_c)$ is given in Fig.~\ref{Jacobi}.
The amplitude-function $\phi_{n\Lambda M_\Lambda}^{(c)}$,
with $M_\Lambda$ the projection of $\Lambda$ on the $z$-axis,
is expanded in terms of the Gaussian basis functions:
\begin{eqnarray}
\phi_{n\Lambda M_\Lambda}^{(c)}({\bf y}_c,{\bf r}_c)
 &=&\sum_{\lambda\ell}\sum_{i=1}^{i_{\rm max}}\sum_{j=1}^{j_{\rm max}}
  A_{n,i\lambda j\ell \Lambda}^{(c)}
  y_c^{\lambda}r_c^{\ell}
  e^{-({y_c}/{\bar{y}_{i}})^2}
  \nonumber \\
 &&\hspace{-5mm}\times{}
  e^{-({r_c}/{\bar{r}_{j}})^2}
   \left[
    Y_{\lambda}(\Omega_{y_c})
    \otimes Y_{\ell}(\Omega_{r_c})
  \right]_{\Lambda M_\Lambda},\label{gauex}
\end{eqnarray}
where $\lambda$ ($\ell$) is the angular momentum
regarding ${\bf y}_c$ (${\bf r}_c$).
The Gaussian range
parameters are taken to be a geometric progression:
\begin{eqnarray}
 \bar{y}_{i}&=&\bar{y}_1
  (\bar{y}_{\rm max}/\bar{y}_1)^{(i-1)/(i_{\rm max}-1)},
 \label{range1}
  \\
 \bar{r}_{j}&=&\bar{r}_1 (\bar{r}_{\rm max}/
  \bar{r}_1)^{(j-1)/(j_{\rm max}-1)}.
 \label{range2}
\end{eqnarray}
The $\Phi_{\gamma}$ is
antisymmetrized for the exchange between $n_1$ and $n_2$; we then have
$A_{n,i\lambda j\ell\Lambda}^{({2})}=
(-)^SA_{n,i\lambda j\ell\Lambda}^{({1})}$,
and $(-)^{\lambda+S}$ must be 1 for $c=3$.
Meanwhile, the exchange between each valence neutron and each nucleon
in $^4$He is treated approximately by the orthogonality condition
model~\cite{OCM}. The eigenenergies $\epsilon_{nI}$ of $^6$He
and the corresponding expansion-coefficients
$A_{n,i\lambda j\ell\Lambda}^{({c})}$ are determined by diagonalizing
$H_6$~\cite{GEM6He1,GEM6He}.
\begin{table*}[htbp]
 \caption{
 The number of the Gaussian basis functions, $i_{\rm max}^{I^\pi(c)}$
 and $j_{\rm max}^{I^\pi(c)}$ for set I, II and III.
 The corresponding number of the eigenstates of $H_6$,
 ${\cal N}_{\rm max}^{I^\pi}$, and the number of channels included
 in the CDCC equation, $N_{\rm max}^{I^\pi}$, are also shown (see the text
 for the details).
 }
 \begin{tabular}{lccccccccccccc}
  \hline
     & & & & & & & & & & 18 MeV &  18 MeV & 229.8 MeV & 229.8 MeV \\
     & $i_{\rm max}^{0^+(3)}$ & $j_{\rm max}^{0^+(3)}$
     & $i_{\rm max}^{0^+(1,2)}$ & $j_{\rm max}^{0^+(1,2)}$
     & ${\cal N}_{\rm max}^{0^+}$
     & $i_{\rm max}^{2^+(1,2,3)}$ & $j_{\rm max}^{2^+(1,2,3)}$
     & ${\cal N}_{\rm max}^{2^+}$
     & $\quad$
     & $N_{\rm max}^{0^+}$ & $N_{\rm max}^{2^+}$
     & $N_{\rm max}^{0^+}$ & $N_{\rm max}^{2^+}$
  \\\hline\hline
  set I   &  8 & 6  &  6 &  6 & 204 &  6 &  6 & 288
          & & 17 & 21 & 28 & 39
  \\
  set II  & 10 & 8  &  8 &  8 & 352 &  8 &  8 & 512
          & & 25 & 32 & 44 & 64
  \\
  set III & 12 & 10 & 10 & 10 & 540 & 10 & 10 & 800
          & & 32 & 42 & 60 & 85
  \\\hline
 \end{tabular}
\label{tab2}
\end{table*}

In the four-body CDCC calculation shown below, we take
$I^\pi=0^+$ and $2^+$ states for $^6$He; $\pi$ is the parity of $^6$He.
We show in Table~\ref{tab1} the maximum values of the internal angular
momenta, $\lambda_{\rm max}$, $\ell_{\rm max}$ and $\Lambda_{\rm max}$,
and the Gaussian range parameters,
$\bar{y}_1$, $\bar{y}_{\rm max}$, $\bar{r}_1$ and $\bar{r}_{\rm max}$,
used in the calculation of $\Phi_{\gamma}$.
It should be noted that most of them depend on $I^\pi$ and $c$, while in
Eqs.~(\ref{gauex})--(\ref{range2}) the dependence has been omitted for
simplicity.
In order to demonstrate the convergence of the four-body CDCC solution
with respect to increasing the number of the Gaussian basis functions,
we prepare three sets of the basis functions, i.e., sets I, II and
III. Each set is specified by $i_{\rm max}^{I^\pi(c)}$ and
$j_{\rm max}^{I^\pi(c)}$; again, the $I^\pi$- and $(c)$-dependence of
them has been omitted in Eqs.~(\ref{gauex})--(\ref{range2}).
One can calculate the total number of the
eigenenergies of $H_6$, ${\cal N}_{\rm max}^{I^\pi}$,
by using Eqs.~(\ref{phig})--(\ref{range2}) and the input
parameters shown in Table~\ref{tab1}.
The values of $i_{\rm max}^{I^\pi(c)}$, $j_{\rm max}^{I^\pi(c)}$ and
${\cal N}_{\rm max}^{I^\pi}$ for each set are shown in Table~\ref{tab2}.
In the actual CDCC calculation for $^6$He+$^{12}$C
scattering at 18 MeV (229.8 MeV), high-lying states with
$\epsilon_{nI} > 12$ MeV ($\epsilon_{nI} > 25$ MeV) are found to
give no effect on the elastic and breakup $S$-matrix elements.
Thus, the effective number of the eigenstates of $^6$He,
$N_{\rm max}^{I^\pi}$, is reduced much for each of sets I--III,
as shown in Table~\ref{tab2}.

As for the coupling potentials in the CDCC equation, we adopt the double
folding model~\cite{dfm} as follows:
\begin{eqnarray}
 U_{\zeta^{\prime}\zeta}^{J}(R)
 &=&
 (N_R+iN_I) V_{\zeta^{\prime}\zeta}^{J}(R), \label{coupl}\\
 V_{\zeta^{\prime}\zeta}^{J}(R) &\equiv&
 \langle
 {\cal Y}_{n'I',L'}^{JM}
 \Phi_{\rm g.s.}^{\rm (T)}
 |
 \sum_{i\in{\rm P}}\sum_{j\in{\rm T}}v_{ij}
 |
 \Phi_{\rm g.s.}^{\rm (T)}
 {\cal Y}_{nI,L}^{JM}
 \rangle
 \nonumber \\
&=&
 \int\rho_{\zeta^{\prime}\zeta}^{{\rm (P)}JM}
 ({\bf r}_{\rm P},\Omega_R)
 \rho_{\rm g.s.}^{\rm (T)}({\bf r}_{\rm T})
 \nonumber \\
 &&\times{}
  v_{\rm NN}(E,\rho,s)\, d{\bf r}_{\rm T}d{\bf r}_{\rm P}d\Omega_R,
\end{eqnarray}
where ${\bf r}_{\rm P}$ (${\bf r}_{\rm T}$) is the coordinate of a
nucleon in the projectile (target) relative to the center-of-mass of
the particle, and
${\bf s}={\bf R}+{\bf r}_{\rm T}-{\bf r}_{\rm P}$.
The quantum number $\zeta$ represents $n$, $I$ and $L$ together, and the
elastic channel, which has the incident wave, is denoted by
$\zeta_0 \equiv (0,I_0,L_0)$.
The ground state density of $^{12}$C,
$\rho_{\rm g.s.}^{\rm (T)}({\bf r}_{\rm T})\equiv\langle\Phi_{\rm
g.s.}^{\rm (T)}| 
\sum_{j=1}^{12}\delta({\bf r}_{\rm T}-{\bf r}_j)|
\Phi_{\rm g.s.}^{\rm (T)}\rangle$, 
where $\Phi_{\rm g.s.}^{\rm (T)}$ is the wave function of $^{12}$C in
the ground state, is calculated by the  microscopic 3$\alpha$ cluster
model~\cite{three-alpha}. 
We in this study define the transition densities of $^6$He,
$\rho_{\zeta^{\prime}\zeta}^{{\rm (P)}JM}$, as
\begin{eqnarray}
 \rho_{\zeta^{\prime}\zeta}^{{\rm (P)}JM}({\bf r}_{\rm P},\Omega_R)=
  \langle
  {\cal Y}_{n'I',L'}^{JM}
  |
  \textstyle{\sum_{i=1}^6}\delta({\bf r}_{\rm P}-{\bf r}_i)
  |
  {\cal Y}_{nI,L}^{JM}
  \rangle_\xi.
\end{eqnarray}
As for the nucleon-nucleon effective interaction $v_{\rm NN}$,
we use the realistic energy- and density-dependent M3Y (DDM3Y)
interaction~\cite{DDM3Y}.
Since the DDM3Y interaction is real,
$V_{\zeta^{\prime}\zeta}^{J}(R)$ has no imaginary part.
Thus, we have multiplied $V_{\zeta^{\prime}\zeta}^{J}(R)$ by a
complex factor $N_R+iN_I$. 
In the present analysis, we fix $N_R=1$ and optimize $N_I$ to fit
experimental data for elastic scattering.
It should be noted that
in three-body CDCC calculation made before for $^6$Li scattering on
various target nuclei~\cite{cdcc1,cdcc-dfm1}, the prescription above
was successful in reproducing experimental data.

\begin{figure}[htb]
\includegraphics[width=0.3\textwidth,clip]{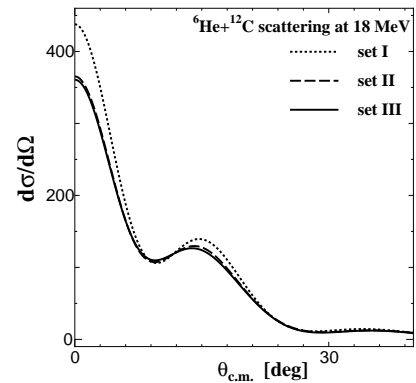}
\caption{Angular distribution of the energy-integrated breakup cross
 section for $^6$He+$^{12}$C scattering at 18 MeV. The dotted, dashed
 and solid lines are the results of the four-body CDCC calculation with
 set I, II and III, respectively, of the Gaussian basis functions.}
\label{break1}
\end{figure}
\begin{figure}[htb]
\includegraphics[width=0.3\textwidth,clip]{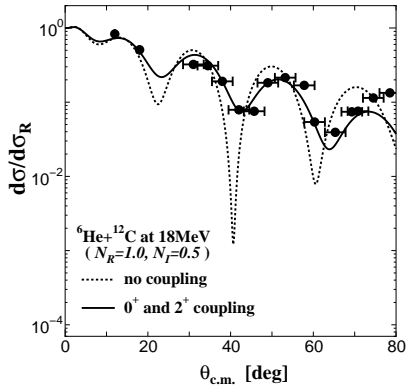}
\caption{Angular distribution of the elastic differential cross section
 for $^6$He+$^{12}$C scattering at
 18 MeV. The solid and dotted lines show the results with and
 without breakup effects, respectively. The experimental data are
 taken from Ref.~\cite{Milin}.}
\label{elastic1}
\end{figure}
\begin{figure}[htb]
\includegraphics[width=0.3\textwidth,clip]{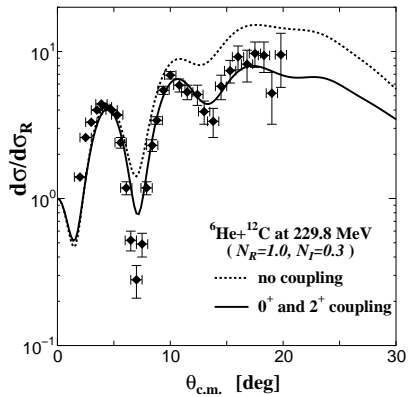}
\caption{The same as in Fig.~\ref{elastic1} but for $^6$He+$^{12}$C
 scattering at 229.8 MeV. The experimental data are taken from
 Ref.~\cite{exp6He1}.}
\label{elastic2}
\end{figure}

The convergence of the four-body CDCC solution is tested for
$^6$He+$^{12}$C scattering at 18 MeV.
Figure \ref{break1} shows the energy-integrated breakup cross section,
i.e., the sum of the cross sections to all breakup channels,
calculated with sets I--III.
The results of set II and set III are
in good agreement with each other, but the result of set I
is somewhat different from them.
Meanwhile, as for the elastic cross section shown in Fig.~\ref{elastic1},
the three sets
give the same cross section shown by the solid line.
Thus, the four-body CDCC solution converges with set II.
Furthermore, we have confirmed that similar convergence is also seen
with respect to extending $\bar{y}_{\rm max}$ and $\bar{r}_{\rm max}$.
The optimum value of $N_I$ determined from the measured elastic cross
section is $0.5$ at $E_{\rm in}=18$ MeV,
which is the same as that for $^6$Li scattering at various
$E_{\rm in}$~\cite{cdcc1,cdcc-dfm1}.
It should be noted that all calculations shown in
Figs.~\ref{break1} and \ref{elastic1} use the same value of $N_I$.
Also for $^6$He+$^{12}$C scattering at 229.8 MeV,
we can see similar convergence
of the elastic and energy-integrated breakup cross sections with respect
to extending the modelspace.
Comparison between the calculated and measured
elastic cross sections is shown in Fig.~\ref{elastic2}.
In this case the optimum value of $N_I$ is 0.3.
In Figs.~\ref{elastic1} and \ref{elastic2},
the dotted lines represent the elastic
cross sections due to the single-channel calculation. Then, the difference
between the solid and dotted lines shows the effect of
the four-body breakup on the elastic cross section.
For both $E_{\rm in}$, the effect is sizable,
properties of which are discussed later.

\begin{figure}[htb]
\includegraphics[width=0.3\textwidth,clip]{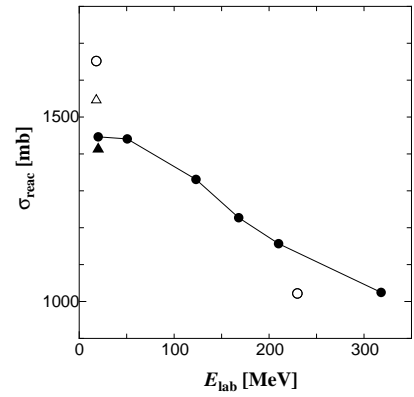}
 \caption{The incident-energy dependence of the total reaction cross
 section for scattering of $^6$He and $^6$Li on  $^{12}$C.
 The open circles show the results for $^6$He+$^{12}$C scattering
 at 18 and 229.8 MeV calculated by four-body CDCC, while the filled
 circles represent those for $^6$Li+$^{12}$C scattering at several energies
 calculated by three-body CDCC based on the $d$+$^4$He
 model for $^6$Li structure.
 The open triangle is the result for $^6$He+$^{12}$C at 18 MeV
 calculated by three-body CDCC with the di-neutron model for
 $^6$He structure.
 The filled triangle is based on the same calculation as
 the open triangle, except that the Coulomb
 potential between $^6$He and $^{12}$C is replaced artificially
 by that between $^{6}$Li and $^{12}$C.
 }
\label{reac}
\end{figure}

Recently, it was reported in Ref.~\cite{dicdc} that the total reaction
cross section for $^6$He+$^{209}$Bi is
much larger than that for $^6$Li+$^{209}$Bi at similar
energies relative to the Coulomb barrier energies because of the large $E1$
excitation strength of $^6$He to the continuum.
Meanwhile, for $^6$He+$^{12}$C scattering at 18 MeV,
the E1 excitation of $^6$He is negligible because $E_{\rm in}$ is
much higher than the Coulomb barrier energy (about 3 MeV).
As shown in Fig.~\ref{reac}, however, we find that 10\% enhancement
of the total reaction cross section is still left.
The open circles represent the total reaction cross sections for
$^6$He+$^{12}$C at 18 and 229.8 MeV calculated by four-body
CDCC, while the filled circles show those for
$^6$Li+$^{12}$C in the energy range 20--318 MeV
calculated by three-body CDCC~\cite{cdcc1,cdcc-dfm1}, where
the microscopic $d$+$^4$He model is assumed for $^6$Li structure.
As mentioned above, the resulting optimum $N_I$ value for
$^6$Li+$^{12}$C scattering is about 0.5,
i.e., almost independent of $E_{\rm in}$.

In order to investigate the origin of the 10\% enhancement, we perform
the three-body CDCC calculation by assuming the di-neutron model for $^6$He
structure; in the model, the di-neutron density is assumed to be the same as
that of the deuteron, and then the resulting $^6$He density is close to
the $^6$Li one. The result of this calculation is shown by the
open triangle in Fig.~\ref{reac}.
The difference between the open triangle and the open circle
at 18 MeV
is due to the Borromean structure of
$^6$He, which is referred to as the Borromean effect. The effect
dominates about half the 10\% enhancement.
The rest of the enhancement is mainly due to the difference of the
Coulomb barrier energies between $^6$He+$^{12}$C and $^6$Li+$^{12}$C.
Actually, when the Coulomb potential for $^6$He+$^{12}$C
is replaced artificially by that for $^{6}$Li+$^{12}$C,
the CDCC calculation based on the di-neutron model
(the filled triangle) gives the
total reaction cross section close to the filled circle at 20 MeV.

As for $^6$He+$^{12}$C scattering at 229.8 MeV,
we have confirmed through the same analysis that
the Borromean effect becomes negligible
as well as the effect of the difference of the Coulomb barrier
between $^6$He+$^{12}$C and $^6$Li+$^{12}$C.
This suggests no enhancement theoretically.
Nevertheless, Fig.~\ref{reac} shows that
the total reaction cross section for $^6$He+$^{12}$C
is even smaller than that for $^6$Li+$^{12}$C
at the similar energy.
This curious behavior is due to the fact
that $N_I=0.3$ for $^6$He+$^{12}$C
while $N_I=0.5$ for $^6$Li+$^{12}$C
at this high energy.
In fact, the total reaction cross section is enhanced by
changing $N_I$ from 0.3 to 0.5 in four-body CDCC calculation
for $^6$He+$^{12}$C, and the resulting cross section almost
reproduces the corresponding one for $^6$Li+$^{12}$C.
The origin of the small
$N_I$ value for the $^6$He scattering is not clear at this moment,
so more systematic
experimental data are highly desirable for $^6$He scattering.

\begin{figure}[htb]
\includegraphics[width=0.3\textwidth,clip]{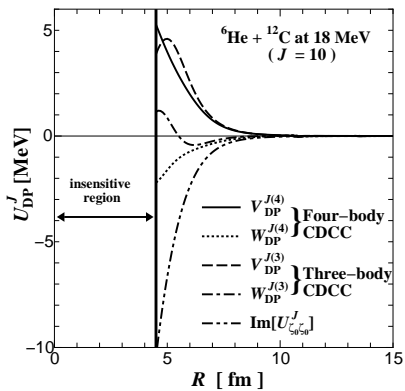}
 \caption{The dynamical polarization potential for
 $^6$He+$^{12}$C scattering at 18 MeV with the grazing angular momentum
 $J_{\rm gr}=10$. The solid and dotted lines, respectively, represent
 the real and imaginary parts of the DP potential calculated by four-body
 CDCC. The dashed and dot-dashed lines correspond to those of
 three-body CDCC with the di-neutron model for $^6$He structure.
 The dot-dot-dashed line represents
 the imaginary part of the double-folded potential $U_{\zeta_0 \zeta_0}^J$.
 }
\label{DPP1}
\end{figure}
Finally, we calculate the dynamical polarization (DP)
potential induced by
the four-body breakup processes, in order to understand effects of the
processes on the elastic scattering. The DP potential $U_{\rm DP}^{J}$ is
given by
\begin{equation}
 U_{\rm DP}^{J}(R)=U_{\rm eq}^{J}(R) - U_{\zeta_0 \zeta_0}^J(R),
\end{equation}
\begin{eqnarray}
\left(
\frac{\hbar^2}{2\mu} \frac{d^2}{dR^2}
+ \frac{\hbar^2}{2\mu} \frac{L_0(L_0+1)}{R^2}
+U_{\rm eq}^{J}(R)+V_{\rm C}(R)-E_{\rm in}
\right)
& &
\nonumber \\
& &
\hspace{-40mm}
\times{}
\chi_{0I_0,L_0}^J(P_{0I_0},R)=0,
\label{Ueq}
\end{eqnarray}
where $\chi_{0I_0,L_0}^J$ is the scattering wave-function in the
elastic-channel calculated with CDCC, and $U_{\rm eq}^{J}$ defined
by Eq.~(\ref{Ueq}) is so-called the equivalent local potential.
The detailed definition of the DP potential is shown in Ref.~\cite{cdcc1}.
Figure \ref{DPP1} shows the DP potential for the $^6$He+$^{12}$C scattering
at 18 MeV with the total grazing angular momentum
$J_{\rm gr}=10$.
The {\lq\lq}insensitive region'' of $R$ shown in the figure
is defined with the
condition that $|\chi_{0I_0,L_0}^J(P_{0I_0},R)|$ is less than 5\% of
its maximum value in the asymptotic region.
The DP potential is almost independent
of $J$ around $J_{\rm gr}$ in the peripheral region.
In Fig.~\ref{DPP1} the real part $V_{\rm DP}^{J(4)}$
($V_{\rm DP}^{J(3)}$) and
the imaginary part $W_{\rm DP}^{J(4)}$ ($W_{\rm DP}^{J(3)}$)
of $U_{\rm DP}^{J}$ calculated by four-body (three-body) CDCC are,
respectively, shown by the solid (dashed) and the dotted (dot-dashed)
lines.
Both of $V_{\rm DP}^{J(4)}$ and $V_{\rm DP}^{J(3)}$
are repulsive and have
almost the same strength which is about 30\% of the real part of
$U_{\zeta_0 \zeta_0}^J$.
The $W_{\rm DP}^{J(4)}$ is about 20\% of the
imaginary part of $U_{\zeta_0 \zeta_0}^J$ (dot-dot-dashed line),
while $W_{\rm DP}^{J(3)}$
oscillates with $R$, so the net effect of $W_{\rm DP}^{J(3)}$
is negligibly small.
Thus, one sees that inclusion of the four-body breakup processes,
i.e., beyond the three-body breakup, makes the real part of
$U_{\zeta_0 \zeta_0}^J$ slightly shallow and the imaginary one deep.
In particular, the latter effect is important and can be assumed
to come from the Borromean structure of
$^6$He. This is consistent with the fact that the total reaction cross
section is enhanced by the Borromean structure.


In conclusion, a fully quantum-mechanical method of treating four-body
nuclear breakup is presented by extending the continuum-discretized
coupled-channels method.
The method called four-body CDCC is applied to $^6$He+$^{12}$C scattering
at 18 and 229.8 MeV in which $^6$He easily breaks up into two neutrons
and $^4$He.
In four-body CDCC, three-body continuum states of $^6$He are discretized by
diagonalizing the internal Hamiltonian of $^6$He with the Gaussian basis
functions.
The validity of four-body CDCC is confirmed by clear convergence of
the calculated elastic and energy-integrated breakup cross sections with
respect to increasing the number of the Gaussian basis functions.
We can say from the convergence that the Gaussian basis functions
form an approximate complete set in a finite configuration
space being important for four-body nuclear breakup processes.
Furthermore, we find a 10\% enhancement of the total reaction cross
section of $^6$He+$^{12}$C at 18 MeV relative to that of
$^6$Li+$^{12}$C at the similar energy.
Half of the 10\% enhancement is due to the Borromean structure of
$^6$He.
For the elastic scattering, the four-body breakup processes
make, in particular, the imaginary part of the double-folded potential deep,
which is originated in
the Borromean structure of $^6$He.
In the present analysis  four-body Coulomb breakup is neglected.
However, it would be possible to treat the Coulomb breakup within the present
framework, if the complex-range Gaussian basis functions are
taken~\cite{Egami}. Further work along this line is highly expected.

The authors would like to thank Y. Sakuragi and M. Kawai for helpful
discussions. This work has been supported in part by the Grants-in-Aid
for Scientific Research (12047233, 14540271) of Monbukagakusyou of
Japan. Numerical calculations were performed on FUJITSU VPP5000
at JAERI.



\begin{thebibliography}{00}

\bibitem{Tanihata}
I. Tanihata {\it et al.},
Phys. Lett. {\bf 160B}, 380 (1985).

\bibitem{adiabatic1}
N. C. Summers {\it et al.},
Phys. Rev. C {\bf 66}, 014614 (2002).

\bibitem{adiabatic2}
J. A. Christley {\it et al.},
Nucl. Phys. {\bf A624}, 275 (1997).

\bibitem{eikonal}
J. S. Al-Khalili {\it et al.},
Nucl. Phys. {\bf A581}, 331 (1995).

\bibitem{exp6He2}
J. S. Al-Khalili {\it et al}., Phys. Lett. B {\bf 378}, 45 (1996).

\bibitem{exp6He1}
V. Lapoux {\it et al.},
Phys. Rev. C {\bf 66}, 034608 (2002).

\bibitem{BAI}
B. Abu-Ibrahim and Y. Suzuki,
Phys. Rev. C {\bf 70}, 011603 (2004).

\bibitem{Milin}
M. Milin {\it et al.},
Nucl. Phys. {\bf A730}, 285 (2004).

\bibitem{cdcc}
M. Kamimura {\it et al.},
Prog. Theor. Phys. Suppl. {\bf 89}, 1 (1986).

\bibitem{CDCC-foundation}
N. Austern {\it et al.},  Phys. Rev. Lett. {\bf 63},
2649 (1989); Phys. Rev. C {\bf 53}, 314 (1996).

\bibitem{dicdc}
N. Keeley {\it et al.},
Phys.~Rev.~C~\textbf{68},~054601 (2003).

\bibitem{pscdc}
T. Matsumoto {\it et al.},
Phys. Rev. C {\bf 68}, 064607 (2003).

\bibitem{GEM}
For a review, E. Hiyama {\it et al.},
Progress in Particle and Nuclear Physics {\bf 51}, 223 (2003).

\bibitem{Piya}
R. A. D. Piyadasa {\it et al}.,
Phys. Rev. C {\bf 60}, 044611 (1999).

\bibitem{OCM}
S. Saito, Prog. Theor. Phys. {\bf 41}, 705 (1969).

\bibitem{GEM6He1}
S. Funada {\it et al.},
Nucl. Phys. {\bf A575}, 93 (1994).

\bibitem{GEM6He}
E. Hiyama and M. Kamimura,
Nucl. Phys. {\bf A588}, 35 (1995).

\bibitem{dfm}
G. R. Satchler and W. G. Love, Phys. Rep. {\bf 55}, 183 (1979).

\bibitem{three-alpha}
M. Kamimura, Nucl. Phys. {\bf A351}, 456 (1981).

\bibitem{DDM3Y}
A. M. Kobos {\it et al.},
Nucl. Phys. {\bf A384}, 65 (1982).

\bibitem{cdcc1}
Y. Sakuragi {\it et al.},
Prog. Theor. Phys. Suppl. {\bf 89}, 136 (1986);
Y. Sakuragi, Phys. Rev. C {\bf 35}, 2161 (1987).

\bibitem{cdcc-dfm1}
C. Samanta {\it et al.}, J. Phys. G {\bf 23},
1697 (1997).

\bibitem{Egami}
T. Egami {\it et al.},
nucl-th/0405050 (2004).

\end{thebibliography}
\end{document}